\documentstyle[aps,preprint]{revtex}
\input epsf
\tightenlines
\begin{document}

\title{Josephson Coupling through a Quantum Dot}

\author{A. V. Rozhkov$^1$, Daniel P. Arovas$^1$, F. Guinea$^2$}

\address{$^1$
Department of Physics, University of California at San Diego,
La Jolla, CA 92093, USA\\
$^2$Instituto de Ciencia de Materiales, Consejo Superior de Investigaciones
Cient{\'\i}ficas, Cantoblanco, 28049 Madrid, Spain}

\maketitle

\begin{abstract}
We derive, via fourth order perturbation theory, an expression for the
Josephson current through a gated interacting quantum dot.
We analyze our expression for two different models of the
superconductor-dot-superconductor (SDS) system.  When the matrix elements
connecting dot and leads are featureless constants, we compute the Josephson
coupling $J_{\rm c}$ as a function of the gate voltage and Coulomb interaction.
In the limit of a diffusive dot, we compute the probability distribution
$P(J_{\rm c})$ of Josephson couplings.  In both cases, $\pi$ junction behavior
($J_{\rm c}<0$) is possible, and is not simply dependent on the parity of the
dot occupancy.

\end{abstract}

\draft
\hfill\break
Magnetic impurities in Josephson junctions tend to degrade the critical current
\cite{kulik,shiso,bula,glamat,spikiv,rozaro,cleamb}.
This result has been known since the work of Kulik \cite{kulik}, who analyzed
a simple model in which spin-preserving as well as spin-flip hopping processes
(with amplitudes $t$ and $t'$, respectively)  are present.  The  critical
current in such a junction is given by the
expression $I_{\rm c}=I^{\rm AB}_{\rm c}\cdot(|t|^2-|t'|^2)/(|t|^2+|t'|^2)$,
where $I^{\rm AB}_{\rm c}$ is the Ambegaokar-Baratoff critical current.
Since spin-flip tunneling results in a sign change of the Cooper pair singlet
($\uparrow\downarrow$ to $\downarrow\uparrow$), the spin-flip hopping
contribution is negative, and reduces $I_{\rm c}$ \cite{spikiv}.
When $I_{\rm c}<0$, one has a $\pi$ junction, for which the ground state
energy is minimized when the superconducting phase difference is $\delta=\pi$.
A ring containing a single $\pi$ junction will enclose trapped
flux \cite{bula}.
A related effect occurs in Josephson tunneling through a ferromagnetic
layer \cite{buzdin,radovic,buzdin2,tanaka}, and recent experiments on
Nb-Cu$_x$Ni$_{1-x}$-Nb junctions suggest that $I_{\rm c}<0$ states
have been observed \cite{SFS}.

In this paper, we investigate Josephson coupling mediated by a quantum dot 
\cite{foot1}, generalizing the case of a single impurity to a system with
many quantized energy levels.  We derive first a general expression,
within fourth-order perturbation theory,
for the Josephson coupling $J_{\rm c}$ ($I_{\rm c}=2eJ_{\rm c}/\hbar$).
We then consider two models for the tunneling amplitudes $t_{\alpha j}$ from
the leads to the dot.  As we shall see,
in contrast to the single impurity case, the parity of the number of electrons
on the dot, $N_0$, does not uniquely determine the sign of the Josephson
coupling.

It is well-known in the theory of elastic co-tunneling that if the Coulomb
repulsion $U$ is large then the conductance through a dot is independent
of $U$.  Correspondingly, we find the critical current is insensitive to
$U$ in this regime, and furthermore when the system is close to a charge
degeneracy point, the probability distribution $P(J_{\rm c})$ for a disordered
dot has universal properties.

Our Hamiltonian is a sum of three terms:
\begin{eqnarray}
{\cal H}_{\rm sc}&=&\sum_{{\bf k},\alpha}
\pmatrix{\psi^\dagger_{{\bf k}\alpha\uparrow}
&\psi_{-{\bf k}\alpha\downarrow}}
\pmatrix{\varepsilon_{{\bf k}\alpha}-\mu & \Delta_\alpha\cr
\Delta^*_\alpha &\mu-\varepsilon_{{\bf k}\alpha}}
\pmatrix{\psi_{{\bf k}\alpha\uparrow}\cr
\psi^\dagger_{-{\bf k}\alpha\downarrow}}\nonumber\\
{\cal H}_{\rm tun}&=&-\sum_{{\bf k},\alpha,j,\sigma}
({{\cal S}_\alpha}{\cal S}_{\rm d})^{-1/2}
\left(t_{\alpha j} \,\psi^\dagger_{{\bf k}\alpha\sigma} c_{j,\sigma} +
t^*_{\alpha j} \, c^\dagger_{j\sigma} \psi_{{\bf k}\alpha\sigma}\right)
\nonumber\\
{\cal H}_{\rm dot}&=&\sum_{j\sigma}\left(\varepsilon_j-\mu+V\right) 
c_{j\sigma}^{\dagger }c_{j\sigma}+\frac{1}{2} N(N-1) U.
\end{eqnarray}
Here, ${\cal S}_{\alpha,{\rm d}}$ are the areas of the $\alpha$ electrode
and the dot, $V$ is the gate voltage on the dot and $U$ is the Hubbard
interaction on the dot.  Electron states on the dot
are assumed to be disordered by a spin-independent random potential.

We calculate the Josephson current via fourth order perturbation
theory in ${\cal H}_{\rm tun}$.  For this to be applicable both the typical
distance between consecutive energy levels $\delta\varepsilon$
and the gap in the superconductors must be large compared to the broadening of
the individual level due to tunneling:
$\Gamma_\alpha \ll \delta \varepsilon,\Delta_\alpha$, where
$\Gamma_\alpha=\pi\nu_\alpha\langle |t_\alpha|^2\rangle/{\cal S}_{\rm d}$,
where $\nu_\alpha$ is the metallic density of states per unit area in
the $\alpha$ electrode.

We also have to be far enough from charge degeneracy points, where the
gap for charge excitations on the dot vanishes.  We measure $V$ relative
to the charge degeneracy point, 
which is equivalent to setting 
$\mu\equiv\varepsilon_{\frac{1}{2}N_0+1} +U N_0$.

In computing the fourth order correction to the ground state energy,
we only consider terms which depend on the phase difference $\delta$
between the two superconductors.
\begin{eqnarray}
E_{\rm J}(\delta)&=&-4\nu_1\nu_2\Delta_1\Delta_2{\rm Re}\,e^{i\delta}\,
\left\{\sum_{\xi_j > 0\atop\xi_{j'}>0} t_{1 j} t_{1 j'}
t^*_{2 j}t^*_{2 j'}J_{\rm ee}(\xi_j,\xi_{j'})\right.\nonumber\\
&&\left.+\sum_{\xi_k < 0 \atop\xi_{k'}< 0} t_{1 k} t_{1 k'} t^*_{2 k}t^*_{2 k'}
J_{\rm hh}(\xi_k,\xi_{k'})-\sum_{\xi_j > 0 \atop\xi_k < 0}
t_{1 j}t_{1 k}t^*_{2 j}t^*_{2 k} J_{\rm eh}(\xi_j,\xi_{k})\right\}\ ,
\label{JC}
\end{eqnarray}
where $J_{\rm ee}, J_{\rm hh}$ and $J_{\rm eh}$ are given by
\begin{eqnarray}
J(\xi,\xi')&=&\int_0^\infty\!\! d\theta_1\int_0^\infty\!\!
d\theta_2\,{\cal J}(\xi,\xi',\theta_1,\theta_2)
\nonumber\\
{\cal J}_{\rm ee}&=&
\left[(\Delta_1\cosh\theta_1+\xi+V)(\xi+\xi'+2V+U)
(\Delta_2\cosh\theta_2+\xi'+V)\right]^{-1}
\nonumber\\
&&+\left[(\Delta_1\cosh\theta_1+\xi+V)
(\Delta_1\cosh\theta_1+\Delta_2\cosh\theta_2)
(\Delta_2\cosh\theta_2+\xi'+V)\right]^{-1}\nonumber\\
&&+\left[(\Delta_1\cosh\theta_1+\xi+V)(\xi+\xi'+2V+U)
(\Delta_2\cosh\theta_2+\xi+V)\right]^{-1}\nonumber\\
{\cal J}_{\rm hh}&=&
\left[(\Delta_1\cosh\theta_1-\xi-V+U)(-\xi-\xi'-2V+3U)
(\Delta_2\cosh\theta_2-\xi'-V+U)\right]^{-1}
\nonumber\\
&&+\left[(\Delta_1\cosh\theta_1-\xi-V+U)
(\Delta_1\cosh\theta_1+\Delta_2\cosh\theta_2)
(\Delta_2\cosh\theta_2-\xi'-V+U)\right]^{-1}\nonumber\\
&&+\left[(\Delta_1\cosh\theta_1-\xi-V+U)(-\xi-\xi'-2V+3U)
(\Delta_2\cosh\theta_2-\xi-V+U)\right]^{-1}\nonumber\\
{\cal J}_{\rm eh}&=&
\left[(\Delta_1\cosh\theta_1-\xi'-V+U)
(\Delta_1\cosh\theta_1+\Delta_2\cosh\theta_2+\xi-\xi')
(\Delta_2\cosh\theta_2-\xi'-V+U)\right]^{-1}\nonumber\\
&&+\left[(\Delta_1\cosh\theta_1+\xi+V)
(\Delta_1\cosh\theta_1+\Delta_1\cosh\theta_2+\xi-\xi')
(\Delta_2\cosh\theta_2+\xi+V)\right]^{-1}\nonumber\\
&&+\left[(\Delta_1\cosh\theta_1+\xi+V)
(\Delta_1\cosh\theta_1+\Delta_2\cosh\theta_2)
(\Delta_1\cosh\theta_1-\xi'-V+U)\right]^{-1}\nonumber\\
&&+\left[(\Delta_2\cosh\theta_2+\xi+V)
(\Delta_1\cosh\theta_1+\Delta_2\cosh\theta_2)
(\Delta_2\cosh\theta_2-\xi'-V+U)\right]^{-1}\nonumber\\
&&+\left[(\Delta_1\cosh\theta_1+\xi+V)
(\Delta_1\cosh\theta_1+\Delta_2\cosh\theta_2+\xi-\xi')
(\Delta_1\cosh\theta_1-\xi'-V+U)\right]^{-1}\nonumber\\
&&+\left[(\Delta_2\cosh\theta_2+\xi+V)
(\Delta_1\cosh\theta_1+\Delta_2\cosh\theta_2+\xi-\xi')
(\Delta_2\cosh\theta_2-\xi'-V+U)\right]^{-1}\ ,
\label{Jf}
\end{eqnarray}
where $\xi\equiv\varepsilon-\mu$.

To proceed further we have to make some assumptions about the tunneling
amplitudes and energy spectrum of the dot.  Initially, we ignore any structure
to the hopping amplitudes and set $t_{\alpha j}\equiv t_\alpha$. This choice
of the amplitudes is
equivalent to a situation where both leads are connected to the same
point on the surface of the dot.

We
further assume the energy spacings on the dot satisfy $\delta\varepsilon\ll
\Delta_\alpha$.  The summation over states on the dot can then be recast
as an integral.  We obtain $E_{\rm J}=J_{\rm c}(1-\cos\delta)$, with
\begin{eqnarray}
J_{\rm c}&=&{4\over\pi^2}\nu_{\rm d}^2\Gamma_1\Gamma_2\Delta_1\Delta_2
\,\int_0^1\!\!dx\,
\int_0^1\!\!dy\,\Big[B_{\rm e}^2\, J_{\rm ee}(B_{\rm e}x,B_{\rm e}y)\nonumber\\
&&\qquad+B_{\rm h}^2\, J_{\rm hh}(-B_{\rm h}x,-B_{\rm h}y)
-B_{\rm e}B_{\rm h}\, J_{\rm eh}(B_{\rm e}x,-B_{\rm h}y)\Big]\ ,
\label{JCint}
\end{eqnarray}
where $B_{\rm e,h}$ are the distance to the top/bottom of the dot spectrum
from the Fermi level on the dot.

By numerical integration we obtain $J_{\rm c}(V)$, shown in fig. 1a for
different values of $U$.  For small values of $U$ the Josephson coupling
always is positive.  However, above some critical value $U_{\rm c}$ there is
a finite interval of voltages in which $ J_{\rm c}<0 $.  The phase diagram
of the junction is presented in fig. 1b-c.   Note that our approach does
not apply close to the $V=0$ or $V=U$ lines where the Coulomb blockade is
lifted and the dot ground state is degenerate.

{\it Disordered dot --}  When the leads are connected to the dot at two 
different points, it is necessary to account for the effects of disorder.
In this case, we have $t_{\alpha j}=t_\alpha\,\psi_j({\bf R}_\alpha)$,
where $\psi_j({\bf r})$ is the wave function corresponding to energy level
$\varepsilon_j$ on the dot.  These wave functions are functionals
of the disorder potential.  The Josephson coupling is now a random quantity
and we must find its distribution, which is possible since the statistical
properties of the $\psi_j$ are well-studied.

Let us define the dimensionless local density of states,
\begin{eqnarray}
\rho_\omega({\bf R}_1,{\bf R}_2)&=&
{G^{\rm A}_\omega({\bf R}_1,{\bf R}_2)-
G^{\rm R}_\omega({\bf R}_1,{\bf R}_2)\over 2\pi i\nu_{\rm d}}\nonumber\\
&=&\nu_{\rm d}^{-1}\sum_j\psi_j({\bf R}_1)\psi_j^*({\bf R}_2)
\delta(\omega-\varepsilon_j)\ ,
\end{eqnarray}
where $\nu$ is the dot density of states
$\nu_{\rm d}=1/({\cal S}_{\rm d}\delta\varepsilon)$, and $G^{\rm R,A}$ are the
usual retarded/advanced single particle Green's functions.

As it was discussed in \cite{alegla} for 
$\omega$ less then Thouless energy $E_{\rm Th}$ these objects have very simple
disorder-averaged statistical properties:
\begin{eqnarray}
\langle \rho_\omega({\bf R}) \rho_{\omega'}({\bf R'})\rangle&=&
\delta\varepsilon\,\delta(\omega-\omega'),\nonumber\\
\langle \rho_{\omega_1}\ldots \rho_{\omega_{2n}}\rangle&=&
\sum_{\rm all\ pairwise \atop contractions} \langle \rho_{\omega_{i_1}}
\rho_{\omega_{i_2}}\rangle\ldots
\langle \rho_{\omega_{i_{2n-1}}} \rho_{\omega_{i_{2n}}}\rangle,
\end{eqnarray}
implying that the function $\rho_\omega$ is distributed according to the
functional
\begin{equation}
P[\rho_\omega]\propto \exp\left\{-\int_{-\infty}^\infty\!\!
d\omega {\rho_\omega^2\over 2\delta\varepsilon}\right\}\ .
\label{gdist}
\end{equation}

The Josephson energy is then a bilinear functional of $\rho_\omega$:
\begin{eqnarray}
J_{\rm c}[\rho_\omega]&=&4\Delta^2 g_1 g_2
\,\Bigg\{\int_0^{\infty}\int_0^{\infty}
d\omega d\omega' \rho_\omega \rho_{\omega'}
J_{\rm ee}(\omega,\omega')\nonumber\\
&&\qquad+\int_{-\infty}^0\int_{-\infty}^0 d\omega d\omega' \rho_\omega
\rho_{\omega'} J_{\rm hh}(\omega,\omega')
-\int_0^{\infty}\int_{-\infty}^0 d\omega d\omega' \rho_\omega
\rho_{\omega'}J_{\rm eh}(\omega,\omega')\Bigg\}
\end{eqnarray}
where we define dimensionless conductances
$g_\alpha=\nu_\alpha\nu_{\rm d} |t_\alpha|^2$.

Formally, our task is straightforward.  We have to obtain critical current 
distribution function 
$P(J_{\rm c})=\langle\delta(J_{\rm c}-J_{\rm c}[\rho_\omega])\rangle$.
We do that as follows.  First, we find the eigenvalues of the
matrix ${\tilde J}(\omega,\omega')$, where
\begin{equation}
2{\tilde J}(\omega,\omega')=\cases{
J_{\rm ee}(\omega,\omega')
+J_{\rm ee}(\omega',\omega)&if $\omega>0$, $\omega'>0$\cr
-J_{\rm eh}(\omega,\omega')&if $\omega>0$, $\omega'<0$\cr
-J_{\rm eh}(\omega',\omega)&if $\omega<0$, $\omega'>0$\cr
J_{\rm hh}(\omega,\omega')
+J_{\rm hh}(\omega',\omega)&if $\omega<0$, $\omega'<0$.\cr}
\label{Jmatrix}
\end{equation}
The distribution $P(J_{\rm c})$ now has the form
\begin{eqnarray}
P(J_{\rm c})&=&\Big\langle\delta\left(J_{\rm c}-4\Delta^2 g_1 g_2
\,[\rho\cdot{\tilde J}\cdot \rho]\right)\Big\rangle
\nonumber\\
&=&\int\limits_{-\infty}^\infty \frac{du}{2\pi}{\rm e}^{-iuJ_{\rm c}}
\prod_\lambda\left(1-4iu(\delta\varepsilon)\Delta^2
g_1 g_2\, \lambda\right)^{-1/2}\ ,
\label{PJint}
\end{eqnarray}
where the product is over all eigenvalues $\lambda$ of the matrix
${\tilde J}$.  Normalization determines the appropriate branch cut.

If both leads are connected to the same point on the surface of the dot,
then $J_{\rm c}\propto\int\! d\omega\int\! d\omega'\,
{\tilde J}(\omega,\omega')$, which we have previously shown to be negative
for certain $V,U$.  Hence  ${\tilde J}$ is not in general positive-definite,
and we anticipate some finite probability for $\pi$ junction behavior.

Our numerical procedure consists of several steps. We introduce a cut-off $W$
and confine $\omega$ to the interval $-W/2<\omega,\omega'<W/2$. Further,
${\tilde J}(\omega,\omega')$ is put on a uniform frequency grid.
The Josephson coupling now is $J_c=4\Delta^2 g_1 g_2
\sum_{s,s'} {\tilde J}_{ss'}\rho_s \rho_{s'}$ where the $2N\times 2N$ matrix
${\tilde J}_{ss'}=(W/2N)\,{\tilde J}(\omega_s\omega_s')$ and
$\rho_s=(W/2N)^{1/2}\,\rho_{\omega_s}; \langle \rho_s \rho_{s'}\rangle=
\delta\varepsilon\,\delta_{ss'}$.
${\tilde J}_{ss'}$ is diagonalized and its eigenvalues $\lambda_s$ are
substituted into (\ref{PJint}).  To make this integral more suitable for
numerical evaluation a contour transformation is performed:
\begin{equation}
P(J_c)=\int_0^{+\infty} \frac{du}{\pi}\,e^{\mp uJ_c}\prod_{s} \left(
\left|1\mp 4u\delta\varepsilon\Delta^2 g_1 g_2 \lambda_s\right|\right)^{-1/2}
\sin\left( \frac{\pi}{2} \sum_s\Theta(\mp u \lambda_s-1)\right).
\end{equation}
One should pick upper sign for $J_{\rm c}>0$ and lower sign for $J_{\rm c}<0$.

The results of numerical calculation are quite insensitive to $W$ and $N$.
The distribution function $P(J_c)$ is presented on fig. 2. 
It has a shape of asymmetric bell and it is non-zero for $J_c<0$ per
previous discussion.  The total probability of having $\pi$-junction:
$P_\pi=\int_{-\infty}^0 P(J_{\rm c})\,dJ_{\rm c}$, is shown in fig. 3.

{\it Universal Limit -- } Generally the distribution $P(J_{\rm c})$ of
Josephson couplings is dependent on nonuniversal parameters such as $V$
and $U$.  In the limit $U \gg \Delta_1=\Delta_2\equiv\Delta$, though,
there are universal aspects to the distribution.  If the dot is close
to the charge degeneracy point, {\it i.e.\/} $\Gamma\ll V\ll \Delta$ or
$\Gamma\ll U-V\ll \Delta$ then the matrix $\tilde J$ is
independent of both $U$ and $V$.  Restricting our attention to the first
case, we find the matrix $\tilde J$ is still given by (\ref{Jmatrix}) with
$J_{\rm hh}=0$ and
\begin{eqnarray}
J_{\rm ee}&\approx&
\int_0^{\infty} d\theta_1\int_0^{\infty} d\theta_2
\left[\left(\Delta\cosh\theta_1+\omega\right)
\left(\Delta\cosh\theta_1+\Delta\cosh\theta_2\right)
\left(\Delta\cosh\theta_2+\omega'\right)\right]^{-1}
\nonumber\\
J_{\rm eh}&\approx&
\int_0^{\infty} d\theta_1\int_0^{\infty} d\theta_2
\left[\left(\Delta\cosh\theta_1+\omega\right)
\left(\Delta\cosh\theta_1+\Delta\cosh\theta_2+
\omega-\omega'\right)
\left(\Delta\cosh\theta_2+\omega\right)\right]^{-1}\ .\nonumber
\end{eqnarray}
Thus, ${\tilde J}(\omega,\omega')=
F(\omega/\Delta,\omega'/\Delta)/\Delta^3$ where $F(x,x')$ is a
universal function of its arguments.  We then find $P(J_{\rm c})=
J^{-1}_{\rm AB}\,f(J_{\rm c}/J_{\rm AB})$, where $f(x)$ is universal and
$J_{\rm AB}=\pi \langle G\rangle\Delta/\hbar$.
The average conductance through the dot at $V=\Delta$ was calculated
in \cite{avenaz} to be $\langle G\rangle =(2\pi e^2/\hbar)
g_1 g_2\,(\delta\varepsilon/\Delta)$.

There are several consequences of this universality.  First, it means that all
moments of this distribution are proportional to powers of $J_{\rm AB}$,
with universal coefficients.  In particular, the ratio of the RMS and
mean critical currents is $\sqrt{\langle(\Delta J_{\rm c})^2\rangle}
/\langle J_{\rm c}\rangle\simeq 1.59$.
Second, the probability of having a $\pi$-junction is a universal number:
$P_\pi=\int_{-\infty}^0 dx f(x)\simeq 0.19.$

{\it Odd $N_0$ -- }
When the parameters of the dot are chosen in such a way that there is a
single level `0' occupied by only one electron, equation (\ref{JC}) must be
modified.  Consider a Cooper pair tunneling from the left to
right superconductor. If none of two electrons passes through `0' then 
their contribution to $E_J$ is already included in (\ref{JC}). Only those
events where one or both electrons tunnel through `0' will modify the
expression for $E_J$.  We find the corrections are given by
\begin{eqnarray}
\Delta J_{\rm ee}(\omega,\omega')&=&\delta\varepsilon\,\delta(\omega')\,
\left(J_{\rm ee}(\omega,0)-\frac{1}{2} J_{\rm eh}(\omega,0)\right)
\Theta(\omega)\nonumber\\
\Delta J_{\rm hh}(\omega,\omega')&=&\delta\varepsilon\,\delta(\omega')\,
\left(J_{\rm hh}(\omega,0)-\frac{1}{2} J_{\rm eh}(0,\omega)\right)
\Theta(-\omega)\nonumber\\
\Delta J_{\rm eh}(\omega,\omega')&=&(\delta\varepsilon)^2\,\delta(\omega)\,
\delta(\omega')\,J_{\rm hh}(0,0)\ .
\end{eqnarray}
In the case of a single impurity level, we recover previous results
\cite{glamat}.  Overall, this electron increases our chances to get $\pi$
junction.  It is clear, however, that the influence of that electron is small
for metallic dot where $\delta\varepsilon\ll \Delta$.

{\it Discussion -- } Two superconductor-dot-superconductor models have
been considered here.  Assuming that both leads are attached to the 
same point on the dot surface and thus, ignoring any structure
to the hopping matrix elements $t_{\alpha j}$, we find a critical $U$ above
which $J_{\rm c}$ can be driven negative by an appropriate dot gate voltage,
$V$.  As one can see from (\ref{JC}) and (\ref{Jf}) there are six terms 
giving positive
contribution to $J_{\rm c}$ and six terms giving a negative contribution.
They each depend in different ways on the gate voltage $V$, and one
can suppress positive terms by varying $V$.

Another feature of our result is that it is derived for an even
number of electrons on the dot.  In the case of an Anderson impurity,
the sign of $J_{\rm c}$ depends on the parity of the impurity level
occupancy $N_0$ \cite{glamat}.  Rather than being a parity effect,
we offer the following interpretation:
If two electrons of the Cooper pair both tunnel through empty
states or both tunnel through filled states this gives positive
contribution to $J_{\rm c}$ (first and second term of (\ref{JC})).  
To obtain a
negative contribution (last term of (\ref{JC})) one electron has to pass
through filled states and the other through empty states.  Thus, for a
completely filled (empty) Anderson impurity both electrons must pass through
filled (empty) states, yielding a positive $J_{\rm c}$.
For a singly occupied orbital,
one electron always tunnels through empty state while the other tunnels
through filled state, and $J_{\rm c}$ is negative.  Therefore, the sign
and magnitude of $J_{\rm c}$ is determined in part by phase space
considerations.

In the second model we allow leads to be `connected' at different points on
the surface of the dot. As a consequence the disorder has to be treated
properly.  We calculated the distribution function of critical current for
the ensemble of junctions for different parameter values. 
The probability of $\pi$ junction as calculated from this distribution has
very reasonable values to expect that such $\pi$ junction can be found 
experimentally.  It is also predicted that
the distribution function should possesses
a remarkable universal property in the limit $U\gg \Delta$.

There are questions that remain unanswered.  Within the framework of our
first model, in order to have $\pi$ junction the value of $U$ must exceed
some critical value $U_{\rm c}$.  From Fig. 3 one might infer that
$U_{\rm c}>0$ for the second model as well.  It has to noted, however, that
our numerical procedure does not rule out the possibility of exponentially 
small tail for $U\rightarrow 0$.  If this is true then $U_{\rm c}$
corresponds to a crossover rather then a phase transition.

\begin{figure} [!t]
\centering
\leavevmode
\epsfxsize=8cm
\epsfysize=8cm
\epsfbox[18 144 592 718] {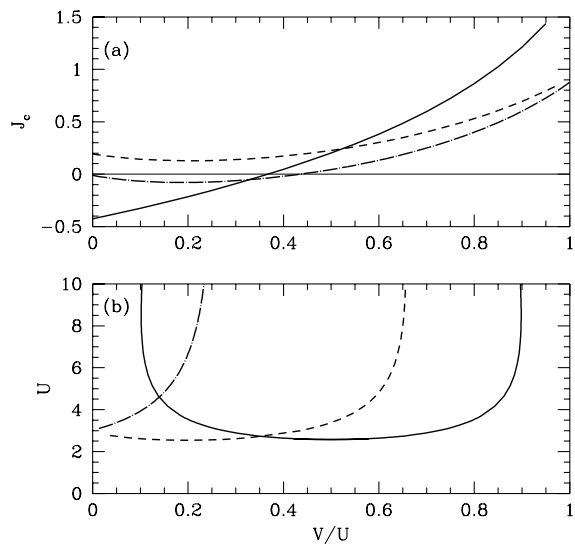}
\caption[]
{\label{fig1} (a) $J_{\rm c}(V)$ {\it versus\/} $V/U$ for $U=5$, $B_{\rm e}=80$,
$B_{\rm h}=20$ (solid); $U=2$, $B_{\rm e}=33.3$, $B_{\rm h}=66.7$ (dashed);
$U=3$, $B_{\rm e}=33.3$, $B_{\rm h}=66.7$ (dot-dashed).  (b) phase diagram
for $B_{\rm e}=50$ (solid),  $B_{\rm e}=33.3$ (dashed), and $B_{\rm e}=10$
(dot-dashed) with $B_{\rm h}=100-B_{\rm e}$  The $\pi$ phase lies above the
curve in each case.}
\end{figure}

\begin{figure} [!h]
\centering
\leavevmode
\epsfxsize=8cm
\epsfysize=8cm
\epsfbox[18 144 592 718] {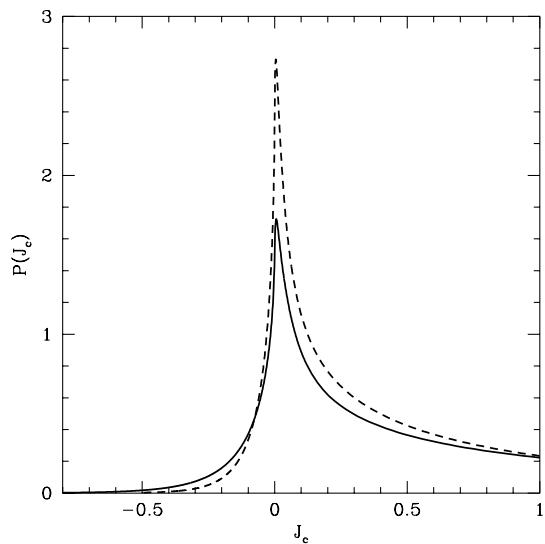}
\caption[]
{\label{fig2} Probability distribution $P(J_{\rm c})$ for $U=3$, $V=0$,
$W=40$ (solid) and $U=3$, $V=1.5$, and $W=40$ (dashed).  $N=100$ in
both cases.  The area under the {\it total} curve ({\it i.e.\/} out to
$|J_{\rm c}|=\infty$ is unity in both cases.}
\end{figure}

\begin{figure} [!h]
\centering
\leavevmode
\epsfxsize=8cm
\epsfysize=8cm
\epsfbox[18 144 592 718] {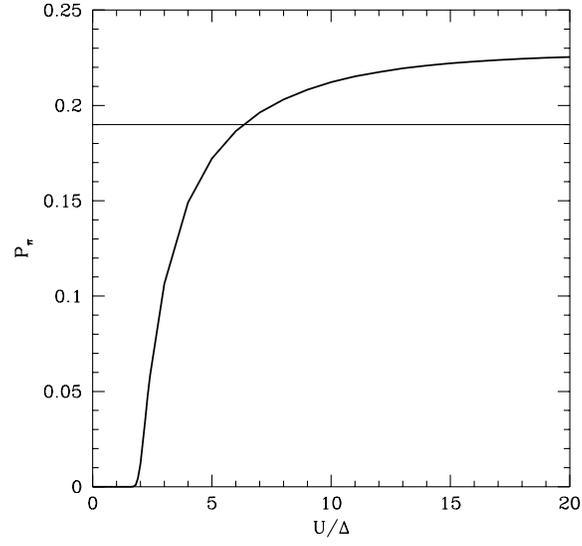}
\caption[]
{\label{fig3} $P_\pi$ {\it versus\/} $U/\Delta$ for $V=0.1\Delta$.
For very large values of $U/\Delta$, $P_\pi\to 0.19$.}
\end{figure}


\begin{thebibliography}{99}

\bibitem{kulik} I. O. Kulik, {\sl Sov. Physics JETP}, {\bf 22}, 841 (1966).

\bibitem{shiso} H. Shiba and T. Soda, {\sl Prog. Theor. Phys.}
{\bf 41}, 25 (1969).

\bibitem{bula} L. N. Bulaevskii, V. V. Kuzii, and A. A. Sobyanin,
{\sl JETP Lett.} {\bf 25}, 290 (1977)

\bibitem{glamat} L. I. Glazman and K. A. Matveev, {\sl Pis'ma Zh. Eksp. Teor.
Fiz.} {\bf 49}, 570 (1989) [{\sl JETP Lett.} {\bf 49}, 659 (1989)].

\bibitem{spikiv} B. I. Spivak and S. A. Kivelson,
{\sl Phys. Rev. B}{\bf 43}, 3740 (1991).

\bibitem{rozaro} A. V. Rozhkov, Daniel P. Arovas,
{\sl Phys. Rev. Lett.} 82, 2788 (1999).

\bibitem{cleamb} A. A. Clerk and V. Ambegaokar, 
{\sl Phys. Rev. B} {\bf 61}, 9109 (2000).

\bibitem{buzdin} A. I. Buzdin, L. N. Bulaevskii, and S. V. Panyukov,
{\sl JETP Lett} 13, 178 (1981).

\bibitem{radovic} Z. Radovi\'c {\it et al.}, {\sl Phys.Rev. B} 44, 759 (1991).

\bibitem{buzdin2} A. I. Buzdin, M. Yu. Kupriyanov, B. Vuji\^ci\'c,
{\sl Physica C} 185-189, 2025 (1991).

\bibitem{tanaka} Yukio Tanaka, Satoshi Kashiwaya, {\sl Physica C}
274, 357 (1997).

\bibitem{SFS} V. V. Ryazanov {\it et al.}, preprint {\tt cond-mat/0008364}.

\bibitem{foot1} We emphasize that we consider only $s$-wave superconductors.

\bibitem{alegla} I.L. Aleiner and L.I. Glazman, 
{\sl Phys. Rev. Lett.} 77, 2057 (1996).

\bibitem{avenaz} D. V. Averin and Yu. N. Nazarov,
{\sl Phys. Rev. Lett.} 65, 2446 (1990).

\end{thebibliography}
\end{document}